\documentclass[%
 aip,
 amsmath,amssymb,
 reprint,%
]{revtex4-1}
\usepackage{xcolor}
\usepackage{graphicx}
\usepackage{dcolumn}
\usepackage{bm}

\usepackage[utf8]{inputenc}
\usepackage[T1]{fontenc}
\usepackage{mathptmx}
\usepackage{etoolbox}

\usepackage{bm}
\usepackage{amsmath}
\usepackage{amsfonts}

\makeatletter
\def\@email#1#2{%
 \endgroup
 \patchcmd{\titleblock@produce}
  {\frontmatter@RRAPformat}
  {\frontmatter@RRAPformat{\produce@RRAP{*#1\href{mailto:#2}{#2}}}\frontmatter@RRAPformat}
  {}{}
}%
\makeatother
\begin{document}

\preprint{AIP/123-QED}

\title{On the Equivalence of Demagnetization Tensors as Discrete Cell Size Approaches Zero in Three-Dimensional Space}

\author{Hao Liang}
 \affiliation{Vanderbilt University Institute of Imaging Science,
Vanderbilt University Medical Center,
Nashville, TN 37232, USA}
 \affiliation{Department of Radiology and Radiological Sciences,
Vanderbilt University Medical Center,
Nashville, TN 37232, USA}
\author{Xinqiang Yan*}%
 \email{xinqiang.yan@vumc.org}
 \affiliation{Vanderbilt University Institute of Imaging Science,
Vanderbilt University Medical Center,
Nashville, TN 37232, USA}
 \affiliation{Department of Radiology and Radiological Sciences,
Vanderbilt University Medical Center,
Nashville, TN 37232, USA}
\affiliation{ 
Department of Electrical and Computer Engineering,
Vanderbilt University,
Nashville, TN 37232, USA
}%

\date{\today}

\begin{abstract}
The calculation of the demagnetization field is crucial in various disciplines, including magnetic resonance imaging (MRI) and micromagnetics. A standard method involves discretizing the spatial domain into finite difference cells and using demagnetization tensors to compute the field. Different demagnetization tensors can result in contributions from adjacent cells that do not approach zero, nor do their differences, even as the cell size decreases. This work demonstrates that in three-dimensional space, a specific set of magnetization tensors produces the same total demagnetization field as the Cauchy principal value when the cell size approaches zero. Additionally, we provide a lower bound for the convergence speed, validated through numerical experiments.

\end{abstract}

\maketitle

\section{Introduction}
Calculation of magnetic field is important in various fields, including  Magnetic Resonance Imaging (MRI) and micromagnetics.
The solution of macroscopic demagnetization field for open boundary conditions is give by \cite{newell1993generalization,abert2019micromagnetics}
\begin{equation}
    \bm{H}
    =
    -\nabla \psi(\bm{r}),
\end{equation}
\begin{equation}
\label{eq:psi}
\psi(\bm{r})
=
-\frac{1}{4\pi}\int \frac{\nabla'\cdot \bm{M}(\bm{r}')}{|\bm{r}  - \bm{r}'|} {\rm d}^3\bm{r}',
\end{equation}
where $\nabla$ is applied to the functions with respect to $\bm{r}$, and $\nabla'$ is applied to the functions with respect to $\bm{r}'$,
$\psi$ is the scalar potential.
Although the function $1\left/|\bm{r}-\bm{r}'|\right.$ with respect to \(\bm{r}'\) diverges at \(\bm{r}\), the divergence is slow, and the integration of Eq. \eqref{eq:psi} exists as an improper integral.
After some mathematical processing, Eq.~\eqref{eq:psi} can be formally written as
\begin{equation}
\label{eq:Hformal}
\bm{H}(\bm{r})
=
-\frac{1}{4\pi}\int 
\bm{M}(\bm{r}')
\cdot
\nabla \nabla'
\frac{1}{|\bm{r}  - \bm{r}'|}
{\rm d}^3\bm{r}'.
\end{equation}
However, the term $\nabla \nabla'\left({1}/{|\bm{r}  - \bm{r}'|}\right)$ as a function of  $\bm{r}'$ divergent rapidly at $\bm{r}$, and the result of Eq.~\eqref{eq:Hformal} depends on how the singularity is treated.
One way is to use the Cauchy principal value
\begin{equation}
\label{eq:Pricipal}
\bm{H}_\text{p}(\bm{r})
=
-\frac{1}{4\pi}\int_{-S} \bm{M}(\bm{r}')
\cdot
\nabla \nabla'
\frac{1}{|\bm{r}  - \bm{r}'|}
{\rm d}^3\bm{r}',
\end{equation}
where $\int_{-S}$ denotes integration over the outside of the sufficient small sphere.

In numerical methods, one class of methods discretizes the spatial domain into finite cells.
The magnetization of each cell is represented by the value at its center, denoted as $\bm{M}(\bm{r})$. The demagnetization field can be expressed as follows:
\begin{equation}
-\sum_{\bm{r}'\neq \bm{r}} N(\bm{r}'-\bm{r})\cdot\bm{M}(\bm{r}'),
\end{equation}
where $N$ is the demagnetization tensor. 
Typically, in micromagnetics, the demagnetization tensor is defined based on the interaction energy between cells, assuming each cubic cell is uniformly magnetized \cite{van2010accuracy, victora2013simulation, ferrero2021adaptive}.
This approach is referred to as the \emph{uniformly magnetized cube (UMC)} method, and the demagnetization tensor of this approach is denoted as $N_\text{c}$.

Another approach treats each cell as a point dipole located at its center.
This approach is used in both micromagnetics \cite{inami2013three,wysocki2017micromagnetic} and MRI \cite{muller2004numerical,muller2005compensation,shang2022high}
owing to its simplicity of calculation for the demagnetization tensor.
This approach is referred to as the \emph{dipole} method, and the demagnetization tensor of this approach is denoted as $N_\text{d}$.

Besides \emph{UMC} and \emph{dipole} methods, the cell at $\bm{r}'$ can be treated as a uniformly magnetized cube; however, the cell at $\bm{r}$ is treated as a point dipole \cite{jenkinson2004perturbation}.
This method can benefit from averaging $\bm{r}'$ over a cell and keeping the calculation relatively simple.
We refer to this method as \emph{uniformly magnetized cube-dipole (UMCD)} method and its demagnetization tensor as $N_\text{cd}$.


The equivalence of these methods is non-trivial due to the singularity of $\nabla\nabla'\left({1}/|\bm{r}-\bm{r}'|\right)$ at $\bm{r}'=\bm{r}$. The field at $\bm{r}$ from a  volume element at $\bm{r}'$ is given by $\sim (\bm{M}{\rm d}^3\bm{r}') \cdot \nabla\nabla' \left(1/{|\bm{r}-\bm{r}'|}\right)$. The term $\nabla\nabla' \left(1/{|\bm{r}-\bm{r}'|}\right)$ diverges as $r^{-3}$ when $r \to 0$, where $r=|\bm{r}-\bm{r}'|$. The volume element ${\rm d}^3\bm{r}'$ can be expressed as $r^2\sin\theta {\rm d}r{\rm d}\psi{\rm d}\theta$ in spherical coordinates. Consequently, the contribution from the small volume diverges as $r^{-1}$. The integral $\int r^{-1} {\rm d}r$ is divergent at $r = 0$. The integral's value depends on the domain of integration. In other words, the demagnetization field is significantly influenced by the shape of the cavity and the location where the field is being evaluated inside the cavity. 

The Cauchy principal value method utilizes an integration over a domain with a spherical cavity. In the \emph{UMC}, \emph{dipole}, and \emph{UMCD} methods, however, the demagnetization field represents the average field in the cubic cell cavity.
Thus, the equivalence of the Cauchy principal value integral method and the \emph{UMC} method must be examined.

Another reason these methods' equivalence is non-trivial is the approximation in the \emph{dipole} method. While $N_\text{d}$ is a good approximation of $N_\text{c}$ for $\bm{r'}$ far from $\bm{r}$, this is not the case for cells adjacent to the cell at $\bm{r}$. 
The contribution of the cell adjacent to the cell at $\bm{r}$ does not approach zero, nor does the difference between different methods, even as the cell size ($h$) approaches zero. The difference can be more than ten percent, regardless of $h$ \cite{della1986magnetization,a1065775}, 
as shown in Fig.~\ref{fig:demag}.
Even more,
there is a difference between the $N_\text{c}$ and $N_\text{cd}$ \cite{fukushima1998volume}, cell by cell.

To our knowledge, the relationship between these methods is not sufficiently discussed. 
Numerous works implicitly assumed that \emph{UMC}, \emph{dipole}, and \emph{UMCD} can give the correct results, i.e., converge to the same result \cite{van2010accuracy, victora2013simulation, ferrero2021adaptive, inami2013three,wysocki2017micromagnetic, muller2004numerical,muller2005compensation,shang2022high}. However, other works hint that these methods can not, due to significant discrepancy between the demagnetization tensors for individual cells~\cite{della1986magnetization,fukushima1998volume,a1065775, Donahue2007}. Additionally, some works treat the result of the \emph{dipole} method as equivalent to that of the Cauchy principal value \cite{shang2022high, muller2005compensation, muller2004numerical}.

This work aims to prove all these methods result in consistent results at the limit of $h$ approaching zero in three-dimensional space, namely 
\begin{equation}
    \lim_{h \to 0} \bm{H}_\text{d}(\bm{r})
    =
    \lim_{h \to 0} \bm{H}_\text{c}(\bm{r})
    =
    \lim_{h \to 0} \bm{H}_\text{cd}(\bm{r})
    =
    \bm{H}_\text{p}.
\end{equation}
The equivalence is not on a cell-by-cell basis but on the total field for sufficient smooth magnetization.

This article is structured as follows:
Section~\ref{sec:theory} reviews the demagnetization tensors, presents the proof of our statement, provides a lower bound on the convergence speed, and outlines the implementation using FFT.
Section~\ref{sec:num} demonstrates the validation of the statement through numerical experiments.
Section~\ref{sec:disc} discusses the extension and limitation of the current work.
Section~\ref{sec:conc} offers concluding remarks.

\section{Theoretical analysis}
\label{sec:theory}
\subsection{Cauchy principal value}
We decompose $\bm{M}(\bm{r}')$ into two functions, $u^{-S}(\bm{r}') \bm{M}(\bm{r}')$ and $u^{S}(\bm{r}')\bm{M}(\bm{r}')$.
Here, $u^{S}=1$ inside a sufficiently small sphere $S$, and $u^{S}=0$ elsewhere; conversely, $u^{-S}=1$ outside the sphere and $u^{-S}=0$ elsewhere. 
This ensures $u^S(\bm{r})+u^{-S}(\bm{r}) \equiv 1$.
Using this notation, $\int$ always means the integral over the whole space:
\begin{equation}
\footnotesize
\label{eq:Decom}
    \bm{H}
    =
\frac{1}{4\pi}
\nabla
\left(
\int \frac{\nabla'\cdot  u^{-S}(\bm{r}') \bm{M}(\bm{r}')}{|\bm{r}  - \bm{r}'|} {\rm d}^3\bm{r}' 
+\int \frac{\nabla'\cdot  u^{S}(\bm{r}') \bm{M}(\bm{r}')}{|\bm{r}  - \bm{r}'|} {\rm d}^3\bm{r}' 
\right).
\end{equation}
The surface magnetic charge has been implicitly counted in Eq.~\eqref{eq:Decom}.

Since $u^{-S} = 0$ inside of the sphere,
we can avoid the singularity issue in evaluating the first part in Eq. \eqref{eq:Decom}.
The first part in Eq.  \eqref{eq:Decom} can be simplified as follows using the method of integration by parts:
\begin{equation}
\label{eq:abc}
\footnotesize
\frac{1}{4\pi}
    \nabla
\int
\left(
\nabla'\cdot \frac{u^{-S}(\bm{r}') \bm{M}(\bm{r}') }{|\bm{r}  - \bm{r}'|}
-
u^{-S}(\bm{r}')\bm{M}(\bm{r}')\cdot\nabla'\frac{1}{|\bm{r}  - \bm{r}'|}
{\rm d}^3\bm{r}' 
\right)
\end{equation}
The first term in \eqref{eq:abc} is simply zero by the Gauss's law, 
noting that $u^{-S}\bm{M}$ inside the sphere and around infinity is zero, respectively. 
Thus, Eq.~\eqref{eq:abc} becomes
\begin{equation}
\begin{aligned}
&-\frac{1}{4\pi}
\nabla
\int
u^{-S}(\bm{r}')\bm{M}(\bm{r}')\cdot\nabla'\frac{1}{|\bm{r}'  - \bm{r}|}
{\rm d}^3\bm{r}' \\
&=
-\frac{1}{4\pi}
\int_{-S}
\bm{M}(\bm{r}')\cdot\nabla\nabla'\frac{1}{|\bm{r}  - \bm{r}'|}
{\rm d}^3\bm{r}'. 
\end{aligned}
\end{equation}
We identify this as the Cauchy principal value, denoted as $\bm{H}_\text{p}$.
The integration of the second term of the integrands in \eqref{eq:Decom} is a well-known problem, and the result is 
$-\frac{1}{3}\bm{M}$ according to, for example, literature \cite{jackson2021classical}.
The macroscopic field $\bm{H}$ can be decomposed into two parts
\begin{equation}
    \bm{H}
    = -
\frac{1}{3} \bm{M}
+\bm{H}_\text{p}.
\end{equation}
The $\bm{H}_{\rm p}$ is given by
\begin{equation}
    \bm{H}_\text{p}(\bm{r})
    =
-\int_{-S}
N_\text{p}(\bm{r}-\bm{r}')
\cdot
\bm{M}(\bm{r}')
{\rm d}^3\bm{r}', 
\end{equation}
where
\begin{equation}
\label{eq:intG}
\begin{aligned}    
N_\text{p}(\bm{r}-\bm{r}')
&=
\frac{1}{4\pi}
\nabla\nabla' \frac{ 1}{|\bm{r}'  - \bm{r}|}
\\
&=
-\frac{1}{4\pi}\frac{3(\bm{r}'  - \bm{r})(\bm{r}'  - \bm{r}) - |\bm{r}'  - \bm{r}|^2}{|\bm{r}'  - \bm{r}|^5}
\end{aligned}
\end{equation}
is the so-called $3\times 3$ demagnetization tensor.
$(N_\text{p})_{ab}$ represents the field component in direction $a$ at position $\bm{r}$, generated by the magnetization component in direction $b$ of a volume element at position $\bm{r}'$, where $a$ and $b$ are indices for axes.
The integration of Eq.~\eqref{eq:intG} is operational because the integration domain excludes the singularity point of $N_\text{p}$. It is important to note that the Cauchy principal value at $\bm{r}$ is distinct from the macroscopic magnetization by $-\bm{M}/3$. More precisely, it represents the field observed in a spherical cavity generated from the remaining parts of the magnetization.

In MRI, the magnetic field with the so-called sphere Lorentz correction, denoted as $\bm{B}'$ is of interest ~\cite{salomir2003fast}, namely
\begin{equation}
    \bm{B}'
    =
    \bm{B}
    -
    \frac{2}{3}\mu_0 \bm{M}
    = \mu_0\left(\bm{H}+\frac{1}{3}\bm{M}\right).
\end{equation}
Thus we have $\bm{B}' = \mu_0 \bm{H}_\text{p}$,
which is already presented in literatures~\cite{wang2015lorentz, wang2015quantitative}.

\subsection{\emph{UMC} and \emph{UMCD} methods}
As mentioned above, \emph{UMC} refers to the approach 
where each cubic cell is uniformly magnetized (however, the magnetization is not necessarily equal for different cells).
In this method, the field is given by
\begin{equation}
\label{eq:BC}
    \bm{H}_\text{c}(\bm{r})
    =
-\sum_{ \bm{r}' \neq \bm{r}}
N_\text{c} (\bm{r}  - \bm{r}')
\cdot  
\bm{M}(\bm{r}').
\end{equation}
The term $\bm{r}' = \bm{r}$ is excluded from the summation for convenient comparison with other methods.
The value of the demagnetization tensor is defined as the average demagnetization field in a cell centered at $\bm{r}$, which is generated by another uniformly magnetized cubic cell (with a unit magnetization) centered at $\bm{r}'$. 
Specifically, the demagnetization tensor is given by:
\begin{equation}
\label{eq:Nc}
    N_\text{c}(\bm{r}-\bm{r}') = \frac{1}{4\pi h^3} \int_{\text{cell } \bm{r}} {\rm d}^3\bm{r}_2 \int_{\text{cell } \bm{r}'} {\rm d}^3\bm{r}_3 \nabla_{\bm{r}_2} \nabla_{\bm{r}_3} \frac{1}{|\bm{r}_2 - \bm{r}_3|}.
\end{equation}

\begin{figure}
    \centering
    \includegraphics[scale=0.8]{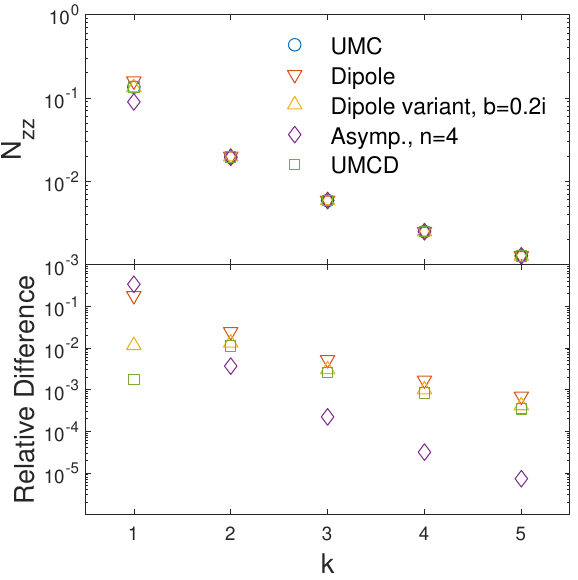}
    \caption{Upper panel: demagnetization tensor, $N_{zz}$, for \emph{UMC} method, \emph{dipole} method, variant of \emph{dipole} method with $b=0.2\imath$, the asymptotic expansion for \emph{UMC} method up to $n=4$, i.e., $O\left(\left(h/r\right)^7\right)$, and \
    \emph{UMCD} method.
    In the plots, $x-x'=y-y'=0$, and $k$ is defined by $z'-z=kh$.    
    Lower Panel: the relative differences between these methods and \emph{UMC} method. Specifically, the relative difference is defined as $\left|{N_{zz}}/{(N_\text{c})_{zz}}-1\right|$ for demagnetization tensor $N$.
    }
    \label{fig:demag}
\end{figure}

The demagnetization tensor $N_\text{c}$ can be calculated using various methods, including numerical integration \cite{van2010accuracy,chernyshenko2015computing}, exact analytical formulas \cite{a1065775,nakatani1989direct, newell1993generalization,fukushima1998volume}, or analytical formulas combined with asymptotic expansions \cite{Donahue2007}.
As an example, $N_{zz}$ is calculated and shown in Fig.~\ref{fig:demag}.

For the \emph{UMCD} method, the cell at $\bm{r}'$ is treated as a uniformly magnetized cube, but the cell at $\bm{r}$ is treated as a point dipole.
This results in a demagnetization tensor of
\begin{equation}
\label{eq:NCD}
    N_\text{cd}(\bm{r}-\bm{r}') = \frac{1}{4\pi} \int_{\text{cell } \bm{r}'} {\rm d}^3\bm{r}_3 \nabla \nabla_{\bm{r}_3} \frac{1}{|\bm{r} - \bm{r}_3|}.
\end{equation}
The analytical expression of the demagnetization tensor for \emph{UMCD}  \cite{jenkinson2004perturbation, fukushima1998volume} is simpler than that for \emph{UMC}.
The $N_\text{zz}$ as an example is shown in Fig.~\ref{fig:demag}. 
Similar to the \emph{UMC} method, $\bm{r}' = \bm{r}$ is excluded from the summation for the \emph{UMCD} method, for convenient comparison with other methods.

However, for both methods,
by noticing that $\text{Tr}\left[\nabla\nabla'\left({1}/{|\bm{r}'-\bm{r}|}\right)\right]
=\nabla\cdot \left(\nabla' \left(1/{|\bm{r}'-\bm{r}|}\right)\right)
=4\pi\delta(\bm{r}'-\bm{r})$ and considering the symmetries of $x$, $y$, and $z$ axes,
it is not difficult to demonstrate that demagnetization tensor at $\bm{r}' = \bm{r}$ is $(1/3)\delta_{ab}$, where 
$\delta{(\bm{r}-\bm{r}')}$ is the Dirac delta function,
$\delta_{ab}$ Kronecker delta,
$\text{Tr}$ denotes the trace of a $3\times 3$ tensor.
Therefore, the demagnetization field generated by the cell itself is $-(1/3)\bm{M}$.
Thus,
\begin{equation}
\bm{H}^\text{full}_\text{c}
\equiv
-\sum
N_\text{c} (\bm{r}  - \bm{r}')
\cdot  
\bm{M}(\bm{r}')
   =\bm{H}_\text{c}(\bm{r})-\frac{1}{3}\bm{M},
\end{equation}
and
\begin{equation}
\bm{H}^\text{full}_\text{cd}
\equiv
-\sum
N_\text{cd} (\bm{r}  - \bm{r}')
\cdot  
\bm{M}(\bm{r}')
=\bm{H}_\text{cd}(\bm{r})-\frac{1}{3}\bm{M}.
\end{equation}

\subsection{\emph{Dipole} method}
As mentioned above, \emph{dipole} method replaces a uniformly magnetized cubic cell with a single dipole at the cell's center.
The demagnetization field is expressed by:
\begin{equation}
\label{eq:BD}
    \bm{H}_\text{d}(\bm{r})
    =
-\sum_{ \bm{r}' \neq \bm{r}}
N_\text{d}(\bm{r}'  - \bm{r})
\cdot
\bm{M}(\bm{r}'),
\end{equation}
where ${N}_\text{d}$ represents the demagnetization tensor for the \emph{dipole} method. 
This tensor quantifies the demagnetization field at $\bm{r}$, generated by a dipole with moment of $h^3$ at $\bm{r}'$, specifically:
\begin{equation}
\label{eq:Gdzz}
{N}_\text{d}(\bm{r}'  - \bm{r})
=
-\frac{3}{4\pi \rho^5}
\begin{pmatrix} 
i^2 -\rho^2/3 & ij & ik \\
ij & j^2 - \rho^2/3 & jk \\
ik & jk &  k^2 - \rho^2/3 \\
\end{pmatrix},
\end{equation}
\begin{equation}
\rho \equiv \sqrt{i^2+j^2+k^2},
\end{equation}
\begin{equation}
\label{eq:rp}
\bm{r}' - \bm{r}
=
(i\bm{e}_x + 
j\bm{e}_y + 
k\bm{e}_z)h.
\end{equation}
where $i$, $j$, and $k$ are integers, and
$\bm{e}_x$, $\bm{e}_y$, and $\bm{e}_z$ are the unit vectors along $x$, $y$, and $z$ directions, respectively.

Note that Eq. \eqref{eq:Gdzz} does not explicitly depend on $h$. Therefore, the contribution from cells adjacent to the cell at $\bm{r}$ does NOT approach zero as $h \to 0$.
It is also noteworthy that since the interaction energy between two uniformly magnetized spheres equals that between two dipoles of the same moment \cite{milton1998classical,newell1993generalization,edwards2017interactions}, the method can also be referred to as the {\it uniformly magnetized sphere} method. Similar to the \emph{UMC} and \emph{UMCD} methods,
$\bm{r}' = \bm{r}$ must be excluded from summation because the demagnetization tensor for the \emph{dipole} method is singular at this point.

\subsection{Asymptotic expansions}
$N_\text{c}$ and $N_\text{cd}$ can be expanded as a series in terms of $1/r$ in the form of
\begin{equation}
    \sum_{n=0,2,...}
    N^{(n)}
=
    \sum_{n=0,2,...}
    \sum_{s,t}
    c_{n,s,t} 
    \left(\frac{x}{r}\right)^s \left(\frac{y}{r}\right)^t
    \left(\frac{z}{r}\right)^{n+2-s-t}
    \left(\frac{h}{r}\right)^{n+3},
\end{equation}
where $c_{n,i,j}$ are coefficients of mathematical constants.
This can also be written in another form:
\begin{equation}
    \sum_{n=0,2,...}
    \sum_{s,t}
    c_{n,s,t} 
    \left(\frac{i}{\rho}\right)^s \left(\frac{j}{\rho}\right)^t
    \left(\frac{k}{\rho}\right)^{n+2-s-t}
    \left(\frac{1}{\rho}\right)^{n+3}.
\end{equation}

Again, $h$ does not explicitly appear in the formula.
For $n=0$, this represents the dipole approximation. For $n=2$, these values are zero for cubic cells,
which explains why the convergence is faster in the \emph{dipole} method compared to the \emph{UMC} and \emph{UMCD} methods.
For $n=4$, the asymptotic expansions for $N_\text{c}$, such as $N_{xx}^{(4)}$ and $N_{xy}^{(4)}$, are given as follows:\begin{equation}
\footnotesize
N_{xx}^{(4)}
=
    \frac{1}{4\pi}
    \cdot
    \frac{7}{16 \rho^{13}}
    \left(
    2 i^6
    -j^6
    -k^6
    -15 i^4(j^2+k^2)
    +15 i^2(j^4+k^4)
    \right),
\end{equation}
and
\begin{equation}
\footnotesize
N_{xy}^{(4)}
=
    \frac{1}{4\pi}
    \cdot
    \frac{7}{16 \rho^{13}}
    ij
    \left(
    7i^4 
    -19 i^2 j^2
    +7j^4
    -13(i^2+j^2)k^2
    +13k^4
    \right).
\end{equation}
For the \emph{UMCD} method,
the correction of $n=4$ to the demagnetization tensor is only half of that for the \emph{UMC} method. 
The asymptotic expansion of the magnetization tensor for the \emph{UMC} is shown in Fig.~\ref{fig:demag}.
The figure shows that the asymptotic expansions for the \emph{UMC} method up to $n = 4$ converge much faster than the dipole approximation. However, higher-order corrections increase the error for the cell adjacent to the cell where the field is being evaluated.

\subsection{Near and far field decomposition}
We use an analysis method similar to that described in literature \cite{jackson2021classical} to split the demagnetization field into near and far fields. The near magnetic field originates from a cubic volume $V$ of size $L \times L \times L$, while the far field encompasses contributions outside this cubic volume.
We can always choose $L$ to be macroscopically small but much larger than $h$, and ensure that $V$ contains exactly $2N+1$ cubic cells, which satisfies:
\begin{equation}
h \ll 1, \quad L \ll 1, \quad \text{and} \quad \frac{L}{h} = 2N+1 \gg 1.
\end{equation}
We will prove that these methods result in the same field under the limit
\begin{equation}
\label{eq:limit}
h \to 0, \quad L \to 0, \quad \text{and} \quad \frac{L}{h} \to +\infty.    
\end{equation}
It is intuitive that these methods yield the same far field, given that $L/h \to \infty$ under the limit defined in equation \eqref{eq:limit}.
Therefore, it is only necessary to prove these methods produce the same near field under the limit \eqref{eq:limit}.

\subsection{Near field for uniform magnetization}
First, we prove that the near part of the demagnetization field produced by all these methods equals zero in the case of uniform magnetization, i.e., the magnetization is uniform in the near cells.
We utilize a similar treatment as in literature \cite{jackson2021classical}.
For the Cauchy principal value method,
the near field can be expressed as
\begin{equation}
    \bm{H}_\text{p}^V (
    \bm{r})
    =
-\int_{V-S}
N_\text{p}(\bm{r}-\bm{r}')
\cdot 
\bm{M}_0 (\bm{r}')
{\rm d}^3\bm{r}' 
.
\end{equation}
The superscript $V$ indicates that only the contribution from the cubic volume $V$ is counted.
We can directly examine that 
$(N_\text{p})_{xx}(\bm{r}-\bm{r}')+(N_\text{p})_{yy}(\bm{r}-\bm{r}')+(N_\text{p})_{zz}(\bm{r}-\bm{r}') = \delta(\bm{r}'-\bm{r})$ according to \eqref{eq:intG}, thus
\begin{equation}
\label{eq:intzz}
\int_{V-S}
\left((N_\text{p})_{xx} + (N_\text{p})_{yy} + (N_\text{p})_{zz}\right)
{\rm d}^3\bm{r}'
=0.
\end{equation}
Since $\bm{r}$ is exactly at the center of $V$, there is the symmetry between $x$, $y$, and $z$ axes.
Thus, the integral of the three terms in \eqref{eq:intzz} are equal and must be equal to zero,
namely
\begin{equation}
\label{eq:intGpzz}
\int_{V-S}
(N_\text{p})_{xx}
{\rm d}^3\bm{r}'
=
\int_{V-S}
(N_\text{p})_{yy}
{\rm d}^3\bm{r}'
=
\int_{V-S}
(N_\text{p})_{zz}
{\rm d}^3\bm{r}'
=0
\end{equation}

For \emph{dipole} method,
we need to evaluate the magnetic from dipoles inside the cube, according to \eqref{eq:BD}, namely
\begin{equation}
\label{eq:BdV}
    \bm{H}_\text{d}^V(\bm{r})
    =
-\sum_{ 
\substack{\bm{r}' \neq \bm{r}\\\bm{r}'\in V}
}
\bm{M}(\bm{r}') \cdot N_\text{d} (\bm{r}  - \bm{r}')
\end{equation}
We can directly examine $(N_\text{d})_{xx}
+(N_\text{d})_{yy}
+(N_\text{d})_{zz} = 0$ according to \eqref{eq:Gdzz}, thus,
\begin{equation}
\label{eq:sumGdzz}
\sum_{\substack{\bm{r}'\in V\\ \bm{r}' \neq \bm{r}}}^{N}
((N_\text{d})_{xx}
+(N_\text{d})_{yy}
+(N_\text{d})_{zz})
=0.
\end{equation}
Since the symmetry between $x$, $y$, and $z$ axis,
summations of the each of three terms in \eqref{eq:sumGdzz} are equal to zero, as follows:
\begin{equation}
\label{eq:eachNzz}
\sum_{ \substack{\bm{r}'\in V\\ \bm{r}' \neq \bm{r}} }
(N_\text{d})_{xx}
=
\sum_{\substack{\bm{r}'\in V\\ \bm{r}' \neq \bm{r}}}
(N_\text{d})_{yy}
=
\sum_{\substack{\bm{r}'\in V\\ \bm{r}' \neq \bm{r}}}
(N_\text{d})_{zz}
=0.
\end{equation}
For the reason of parity (mirror) symmetry,
the off-diagonal elements of the demagnetization tensor are also zero:
\begin{equation}
\label{eq:intGpxy}
\int_{V-S}
(N_\text{p})_{xy}
{\rm d}^3\bm{r}'
=
\int_{V-S}
(N_\text{p})_{yz}
{\rm d}^3\bm{r}'
=
\int_{V-S}
(N_\text{p})_{xz}
{\rm d}^3\bm{r}'
=0,
\end{equation}
\begin{equation}
\label{eq:eachNxy}
\sum_{\substack{\bm{r}'\in V\\ \bm{r}' \neq \bm{r}}}
(N_\text{d})_{xy}
=
\sum_{\substack{\bm{r}'\in V\\ \bm{r}' \neq \bm{r}}}
(N_\text{d})_{yz}
=
\sum_{\substack{\bm{r}'\in V\\ \bm{r}' \neq \bm{r}}}
(N_\text{d})_{xz}
=0.
\end{equation}

For the \emph{UMC}, \emph{UMCD}, and their asymptotic expansions methods,
the symmetries also exist, allowing similar equations to  Eq.~\eqref{eq:eachNzz} and~\eqref{eq:eachNxy} to hold. 
Note that \emph{dipole} method can be seen as the leading order of \emph{UMC} and \emph{UMCD}'s asymptotic expansions.
In summary, we have proved that,
for all these methods,
the near part of the demagnetization field is exactly zero, namely,
\begin{equation}
\label{eq:BdVZero}
    \bm{H}_\text{p}^V(\bm{r})
    =
    \bm{H}_\text{c}^V(\bm{r})
    =
    \bm{H}_\text{cd}^V(\bm{r})
    =
    \bm{H}_\text{d}^V(\bm{r})
    =0.
\end{equation}

Modifications to the demagnetization tensors, while preserving Eq. \eqref{eq:sumGdzz} and symmetries of $x$, $y$, and $z$ axes, can also result in \eqref{eq:eachNzz}. 
This guides us in creating variation methods of the \emph{dipole} method.
An example is replacing $\rho=\sqrt{i^2+j^2+k^2}$ in denominator in Eq. \eqref{eq:Gdzz} with $\rho=\sqrt{i^2+j^2+k^2 + b}$, namely,
\begin{equation}
\label{eq:Gdzzvar}
{N}_\text{d}(\bm{r}'  - \bm{r})
=
-\Re \frac{3}{4\pi \rho'^5}
\begin{pmatrix} 
i^2 -\rho^2/3 & jk & ik \\
jk & j^2 - \rho^2/3 & jk \\
ik & jk &  k^2 - \rho^2/3 \\
\end{pmatrix},
\end{equation}
\begin{equation}
\rho = \sqrt{i^2+j^2+k^2},~
\rho' = \sqrt{i^2+j^2+k^2+b},
\end{equation}
where $b$ can be positive real number or imaginary number, $\Re \star$ denote the real part of $\star$.
The additional term $b$ softens the demagnetization tensor, possibly leading to a smoother result where $\bm{M}$ changes rapidly.

\subsection{Near field for nonuniform magnetization}

In the previous subsection, it has been established that the near demagnetization field is exactly zero for uniform magnetization.
Then, we prove that the near demagnetization field at $\bm{r}$ approaches zero as $L \to 0$, provided that the magnetization is H\"older continuous at $\bm{r}$. 

By subtracting a constant magnetization, $\bm{M}_0 = \bm{M}(\bm{r})$, from $\bm{M}(\bm{r}')$, the result remains unaffected. Consequently, with this subtraction, $\left.(\bm{M}(\bm{r}') - \bm{M}_0)\right|_{\bm{r'} = \bm{r}} = 0$, allowing us to potentially address the challenges raised from the discrepancy of the demagnetization tensor around $\bm{r}$.

For the principal integral method, we have
\begin{equation}
    \bm{H}_\text{p}^V(\bm{r})
    =
-\int_{V-S}
(\bm{M}(\bm{r}') - \bm{M}_0) \cdot N_\text{p}(\bm{r}-\bm{r}'){\rm d}^3\bm{r}' .
\end{equation}
We assume $M(\bm{r}')$ hold H\"{o}lder condition for $\bm{r}$, namely
 \begin{equation}
 \label{eq:cont}
     |\bm{M}(\bm{r}') - \bm{M}_0| < K |\bm{r'}-\bm{r}|^\alpha,
 \end{equation}
 where $K$ is a constant independent of $\bm{r}'$ and $\alpha > 0$.
 
First, we can estimate the magnitude of the demagnetization field for the principal integral method.
Since the $N_\text{p}$ diverges as $r^{-3}$ as $r\to 0$, thus the integral in spherical coordinates is bounded by
\begin{equation}
\label{eq:ContBpV}
    H_\text{p}^V
    \sim
\int_{V-S}
Kr^\alpha
(1/r^3)
r^2 {\rm d} r
{\rm d} \theta
{\rm d} \phi
\sim K L^{\alpha}.
\end{equation}
In the estimation, we neglect constants on the order of one.
Eq. \eqref{eq:ContBpV} approaches zero as $L\to 0$.
Then, we turn to the \emph{dipole} method. Similarly, we can estimate the magnitude of the demagnetization field of the \emph{dipole} method as
\begin{equation}
\label{eq:ContBdV}
\begin{aligned}
    H_\text{d}^{V}(\bm{r})
    &\sim
    \sum_{i,j,k} K (\rho h)^{\alpha} (1/{\rho}^3) \\
    &\sim
    \sum_{I=1}^N \sum_{\max(|i|,|j|,|k|)=I} 
    K (I h)^{\alpha} (1/I^3) \\
    &\sim
    \sum_{I=1}^N  
    K (I h)^{\alpha} I^2/I^3 \\
    &\sim
    \sum_{I=1}^N
    K h^{\alpha} I^{\alpha-1}
    \sim
    K L^\alpha.
\end{aligned}
\end{equation}
Thus, $\bm{H}_\text{d}^{V}(\bm{r}) \to 0$ at limit of $L\to 0$.

The differences between the \emph{dipole} method and other methods are bounded by the higher-order terms in the asymptotic expansions, possibly multiplied by a factor of order one. 
For these higher-order corrections, we replace $\rho^{-3}$ with $\rho^{-(3+n)}$ in Eq.~\eqref{eq:ContBdV}. After summing over all cells in $V$, the higher-order correction is on the order of $KL^{\alpha}(1^{-n}-N^{-n})$, where $N \sim L/h$. Thus, the higher-order correction to the near field is on the same order as that for the \emph{dipole} method.

In summary, provided $\bm{M}(\bm{r}')$ holds H\"{o}lder condition for $\bm{r}$, all these methods result in a zero near demagnetization field as $L \to 0$, specifically
\begin{equation}
\label{eq:BdVZero}
    \lim_{L\to 0} \bm{H}_\text{d}^V(\bm{r})
    =
    \lim_{L\to 0} \bm{H}_\text{c}^V(\bm{r})
    =
    \lim_{L\to 0} \bm{H}_\text{cd}^V(\bm{r})
    =
    \lim_{L\to 0} \bm{H}_\text{p}^V(\bm{r})
    =0.
\end{equation}
Note that to prove the demagnetization field is zero for a sufficiently small cubic volume centered at $\bm{r}$, 
 Eq.~\eqref{eq:cont} is sufficient. It does not necessarily require H\"{o}lder continuity to hold everywhere in $V$.

\subsection{Lower bound on the convergence speed}

We have proved that the near field is zero as $L\to 0$, assuming H\"{o}lder condition holds at $\bm{r}$.
All methods intuitively yield the same result for the far field; thus, a detailed analysis is not provided. 
Now, We give a lower bound on the convergence speed with the specific condition for magnetization, i.e., the H\"{o}lder condition holds for every cell.
All these methods,
including the exact Cauchy principal value method,
yield values on the order of $K L^\alpha$ for the near field,
excluding $\bm{r}'=\bm{r}$ from summation.
Consequently, the error in the near field is of the order
\begin{equation}
    KL^\alpha.
\end{equation}
For the analysis of the far field,
we begin with the \emph{UMCD} method, in which each cell has an error on the order of $K h^\alpha / \rho^3$. 
The total error for the far field  is bounded by the order of
\begin{equation}
\label{eq:farfield}
    \sum_{i,j,k \notin V} \frac{K h^\alpha}{\rho^3}
    \sim
    \sum_{I \notin V} \frac{K h^\alpha I^2}{I^3}
    \sim
    K h^\alpha \log(L^\text{max}/L),
\end{equation}
where $L^\text{max}$ is the maximum material length.
Taking $L \sim h$, the total error of near and far field is on the order of 
\begin{equation}
\label{eq:nearandfar}
K h^\alpha\log(L^\text{max}/h).   
\end{equation}
For the other methods,
the error has two parts: first, the error of the \emph{UMCD} method, and second, the differences between these methods.
For the far field,
the difference between these methods is bounded by the asymptotic expansion of $n=4$, i.e., $|\max M|/\rho^7 $ for a cell. After summing over all cells, the total error is bounded by the order of 
\begin{equation}
\label{eq:farFieldAanother}
|\max M| \left( h/L \right)^4.    
\end{equation}
The total error for the near field and the far field part of Eq.~\eqref{eq:farFieldAanother} is minimized to the order of
\begin{equation}
\label{eq:NearFarAnother}
 h^{4\alpha/(4+\alpha)},
\end{equation}
for $L \sim h^{4/(4+\alpha)}$. 
For $L \sim h^{4/(4+\alpha)}$, the error of far field for 
\emph{UMCD} method as in Eq.~\eqref{eq:farfield} is on the order of $K h^\alpha \log(L^\text{max}/h)$ is smaller than the error described in Eq.~\eqref{eq:NearFarAnother} and can be neglected. 
Given that this analysis is conservative and the errors are over-estimated, 
the actual convergence speed may exceed these estimates.

\subsection{Implementation using FFT}
Noticing the translation symmetry property of the demagnetization tensor, the summation of the demagnetization tensor weighted by magnetization is a convolution operation. According to the cyclic convolution theorem, cyclic convolution can be performed using fast Fourier transform (FFT) and inverse FFT (IFFT) operations. To avoid the side effects of cyclic convolution, zero padding on the original array $M$ is needed. The detailed steps are as follows:
(1) Evaluate the array $N$ of dimension $N_x \times N_y \times N_z$ representing the demagnetization function, assuming $\bm{r}$ is at the center of array.
Then, circularly shift the center to the frontmost position.
(2) Extend the dimension of $M$ and $N$
from $N_x \times N_y \times N_z$
to $(2N_x-1) \times (2N_y-1) \times (2N_z-1)$ by padding zeros on the end side.
(3) Apply the cyclic convolution to $M$ and $N$ according to the cyclic convolution theorem \cite{press1988numerical}:
   \begin{equation}
   M \otimes G = \text{IFFT}(\text{FFT}(M) \times \text{FFT}(N)),
   \end{equation}
   where $\otimes$ is the cyclic convolution operation,
   $\times$ is the element-wise product operation.
(4) Finally, clip the $N_x \times N_y \times N_z$ elements of $M \otimes G$ at the front.
Except for rounding errors, the result would be identical to the direct summation.
However, the time complexity is reduced from $O(N_x^2N_y^2 N_z^2)$ to $O(N_xN_y N_z\log(N_x N_y N_z))$ using FFT.
Noticing the $\text{FFT}(N)$ is independent of $M$,
thus can be reused for different $M$s provided the dimension of the problem is unchanged.

\section{Numerical Validation}
\label{sec:num}
\begin{figure}[!htbp]
    \centering
    \includegraphics[width=0.95\linewidth]{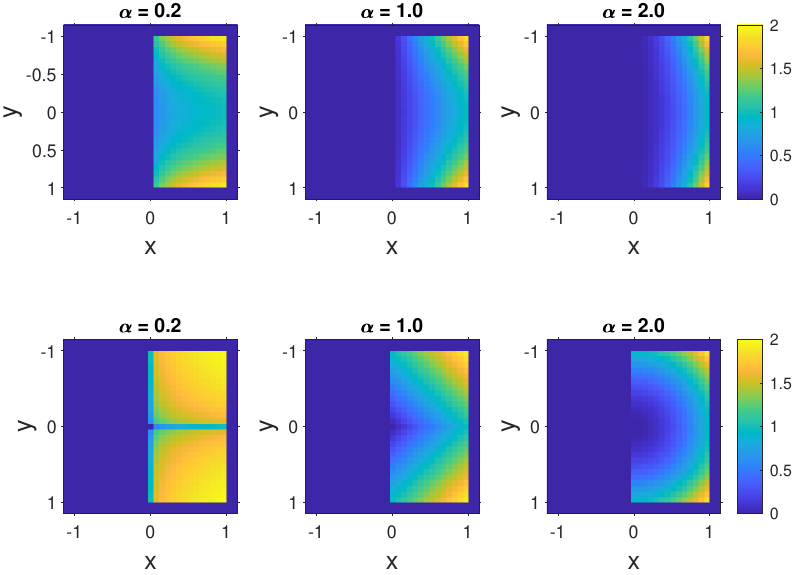}
    \caption{Upper panel: the magnetization for $z=0$ described by Eq. \eqref{eq:magfield1} for different $\alpha$s. Lower panel: the magnetization for $z=0$ described by Eq. \eqref{eq:magfield2} for different $\alpha$s.} 
    \label{fig:mag}
\end{figure}

To validate the conclusion that the three methods agree under certain conditions, we construct two problems. The problem I with magnetization
\begin{equation}
\footnotesize
\label{eq:magfield1}
    \bm{M} = \bm{e}_z\begin{cases} 
    |x|^\alpha(1 + y^2 + z^2) & \text{for $0 \le x \le 1$, $|y|\le 1$, and $|z|\le 1$,} \\
    0 & \text{otherwise,}
    \end{cases}
\end{equation}
and problem II with magnetization
\begin{equation}
\footnotesize
\label{eq:magfield2}
    \bm{M} = \bm{e}_z\begin{cases} 
    |x|^\alpha + |y|^\alpha + |z|^\alpha & \text{for $0 \le x \le 1$, $|y|\le1$, and $|z|\le1$,} \\
    0 & \text{otherwise,}
    \end{cases} 
\end{equation}
where $\alpha > 0$. 
The magnetizations are shown in Fig.~\ref{fig:mag}.
For the problem I, the magnetization is discontinuous at the boundaries of the magnet at $x=1$, $z=\pm 1$, and $y=\pm 1$. While the magnetization is continuous at the boundary of the magnet at $x=0$, its derivative is not for $0 < \alpha \le 1$.
For the problem I, at the center of each cell, $\bm{r}$,
we have $|M(\bm{r}')-M(\bm{r})| \leq 3|\bm{r}'-\bm{r}|^\alpha$,
thus it satifies the H\"{o}lder continuous condition.
For the problem II, the magnetization is discontinuous at the boundaries except for the point $x=y=z=0$.
At $x=y=z=0$, 
the magnetization is continuous; however, its derivative is not for $0 < \alpha \le 1$.
The H\"{o}lder condition still holds at $\bm{r} = 0$.

For the problem I,
the exact analytical result can be expressed using special functions, as shown in  App.~\ref{sec:app1}, with the aid of symbolic processing software \cite{Mathematica}.
Thus, the result can be effectively calculated to arbitrary precision.
We calculate the result to at least 20 digits, although the precision is truncated to about 16 digits when converting from a multi-precision presentation to a double-precision floating-point presentation.
For problem II, the exact result is zero for symmetry reasons.

\begin{figure}[!htbp]
    \centering
    \includegraphics[scale=0.6]{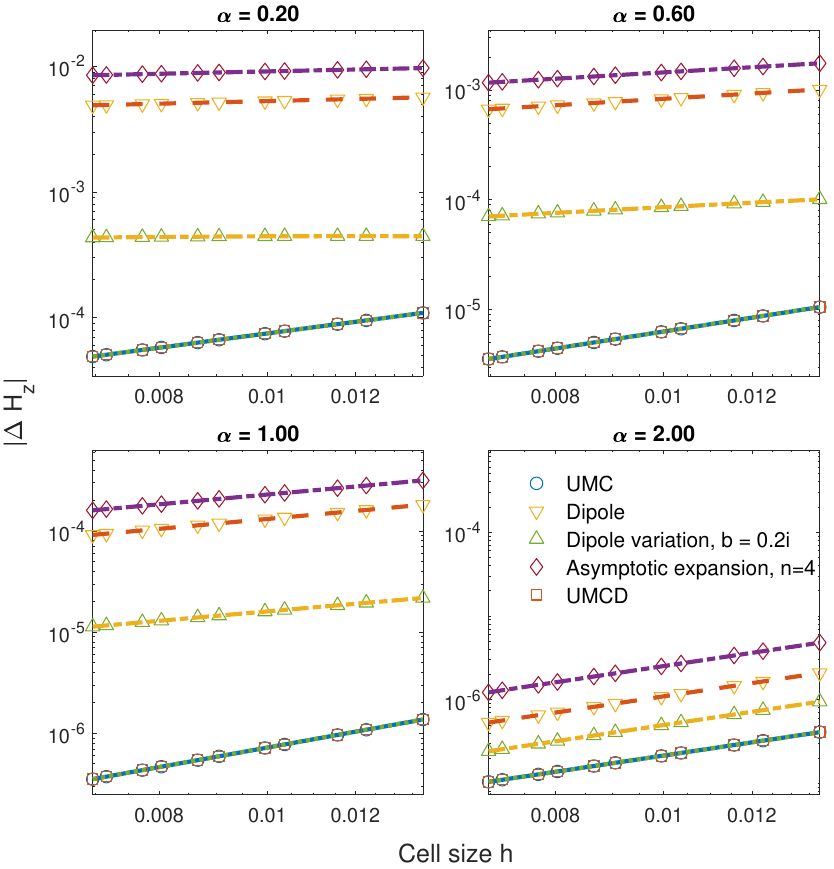}
    \caption{Error in the demagnetization field of numerical calculations as a function of cell size using magnetization for problem I  and for different values of $\alpha$.}
    \label{fig:numerical1}
\end{figure}
\begin{figure}[!htbp]
    \centering
    \includegraphics[scale=0.6]{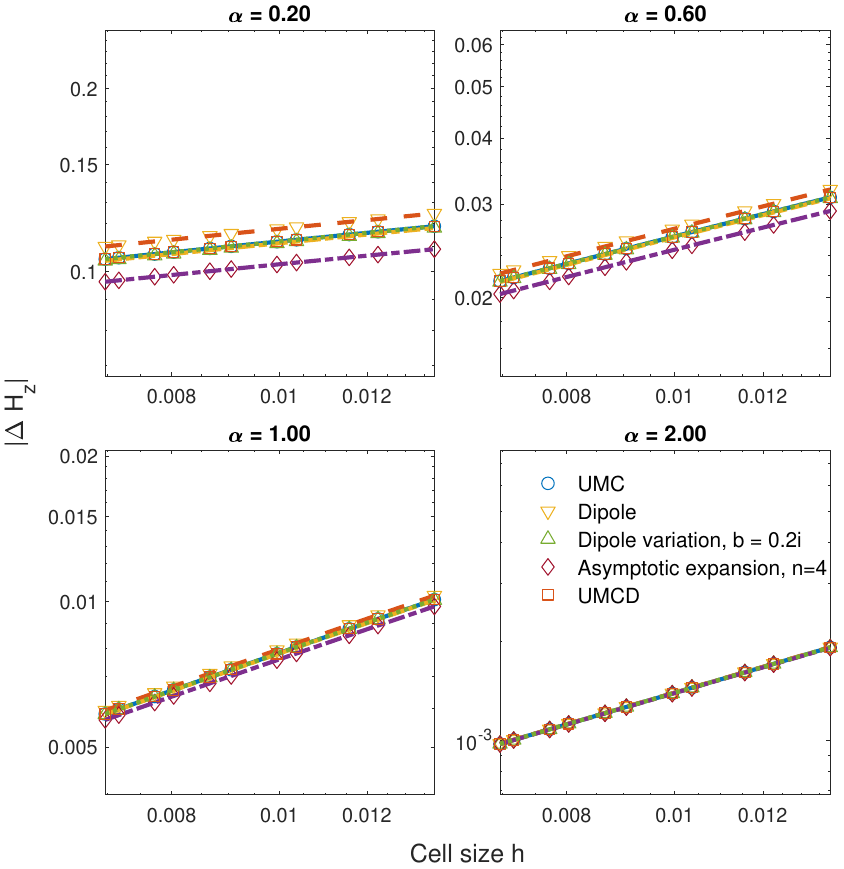}
    \caption{Error in the demagnetization field of numerical calculations as a function of cell size using magnetization for problem II  and for different values of $\alpha$.}
    \label{fig:numerical2}
\end{figure}

\begin{figure}[!htbp]
    \centering
    \includegraphics[width=0.9\linewidth]{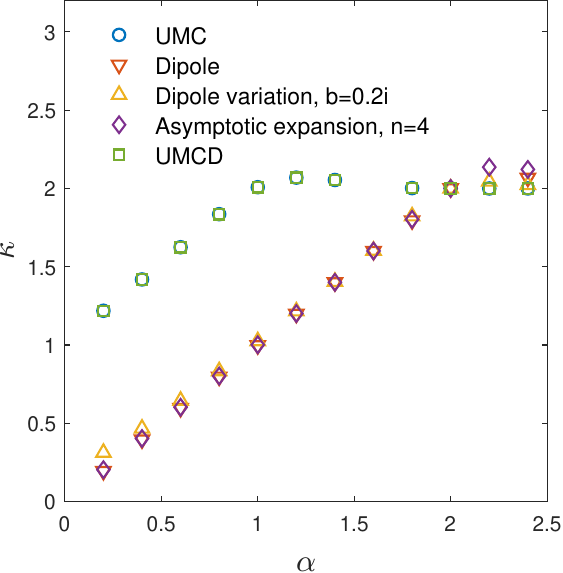}
    \includegraphics[width=0.9\linewidth]{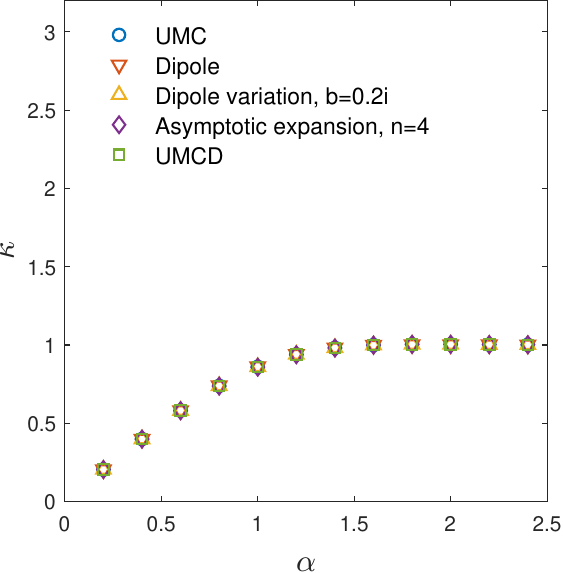}
    \caption{The converge speed as as function of $\alpha$ for problem I (upper panel) and problem II (lower panel).}
    \label{fig:kappavsalpha}
\end{figure}
On the numerical side, we employ several methods: the \emph{UMC} method, the 
\emph{dipole} method, and a variant of the \emph{dipole} method described in $ \eqref{eq:Gdzzvar} $ with $ b = 0.2 \imath $, where $ \imath $ represents the imaginary unit,
the asymptotic expansion for \emph{UMC} and \emph{UMCD} methods.
In the asymptotic expansion method,
the asymptotic expansions are used for both far and near cells.
The coordinate system is set such that the origin is at the center of a cell. The cell size is adjusted so that at $x=1$, $y=\pm 1$, and $z=\pm 1$, the boundaries of the magnet align with the boundaries of the cells. At $x=0$, the cells' centers are on the magnet's boundary.
Thus, we minimize the error fluctuations arising from the discretization of the magnet boundary.
Computations are performed using double-precision floating-point arithmetic. The demagnetization tensors for \emph{UMC} and \emph{UMCD} methods are calculated using exact formulas for cells at short distances and asymptotic expansions for cells at long distances. These asymptotic expansions include terms up to $O\left(\left(h/r\right)^7\right)$. The demagnetization tensor has a maximum error on the order of $10^{-12}$ around the crossover between analytical formulas and asymptotic expansions. The numerical errors observed in this experiment, which are larger than $10^{-7}$, significantly exceed those of double-precision floating-point arithmetic ($10^{-16}$) and the error of the demagnetization tensors. Therefore, all errors are attributed to discretization errors.

The numerical errors of $H_z$ for $\alpha$ of 0.2, 0.6, 1, and 2 at $\bm{r} = 0$ are illustrated in Fig.~\ref{fig:numerical1} and \ref{fig:numerical2}.
The numerical errors of $H_z$ are modeled by $c \cdot h^{\kappa + d (\log{h})^{-2} }$.
The parameter $\kappa$ determines the convergence speed as $h \to 0$.
The $\kappa$ as a function of $\alpha$ are shown in Fig.~\ref{fig:kappavsalpha}.
We observe the $\kappa > 0$ for $\alpha > 0$.
Thus, we verify that these methods converge to the same value for $\alpha > 0$, which is consistent with the prediction of our statement.
All convergence speeds slow down as $\alpha\to 0$ where the magnetization becomes more non-smooth.
In problem I, the convergence speeds for the \emph{UMC} method and \emph{UMCD} method are much faster and have almost identical convergence speeds.
These two methods are significantly better than the other methods. 
This can be explained as follows: for the magnetization relevant to the term $x^\alpha \bm{e}_z$, the volume magnetic charge is zero and only surface magnetic charges appear on $z=\pm 1$.
For these two methods, the error can be attributed to the discretization error for the magnetic surface charge at the magnet surface of $z=\pm 1$. This surface is far from where the field is under evaluation; thus, the error is suppressed. For the magnetization relevant to $x^\alpha (y^2+z^2)\bm{e}_z$, the discretization error from the near field is highly suppressed by the term $x^2+y^2$.
Thus, these two methods show an advantage over the other methods.
We also note that the variation of the \emph{dipole} method shows a better convergence speed than the \emph{dipole} method.
The asymptotic expansion does not take advantage compared to other method, this is as expected, since the asymptotic expansions are not good for near cells.
The convergence speed is faster than our conservative estimate; thus, our results are not violated.
For $\alpha \ge 2$, the $\kappa$ value is close to 2 for all methods, consistent with other studies \cite{miltat2007numerical,abert2013numerical}.
In Problem II, all these methods exhibit similarly slow convergence speeds.
This issue can be attributed to the discontinuity at $x=0$, which worsens the results from all methods.
Additionally, as $\alpha$ approaches zero, the $\kappa$ for all methods approaches zero, indicating non-convergence. This confirms the assertions of this work.

\section{Discussions}
\label{sec:disc}

Some references use $N_{\text{c}}$ for cells near cell to $\bm{r}$ and use the asymptotic approximations for far cells \cite{della1986magnetization}.
Indeed, the summation of demagnetization tensor over any cubic shells, 
$\max(|i|,|j|, |k|)=I$, is zero.
If the crossover happens on a cubic shell, 
\mbox{$\max(|i|,|j|, |k|)=I^\text{crossover}$},
it will maintain similar equations to Eqs.~ \eqref{eq:sumGdzz} and \eqref{eq:eachNxy} and will result in the same limit of demagnetization field as $h\to 0$.
It should be noted that this conclusion does not require $I^\text{crossover}$ to be sufficiently large; it only requires $h \to 0$.

One might use a rectangular prism instead of a simple cube.
For the uniform magnetized prism method, the macroscopic demagnetization field can still be correctly calculated \cite{miltat2007numerical}.
However, in the \emph{dipole} method, using different cell sizes along different directions will disrupt Eqs.~\eqref{eq:sumGdzz} and \eqref{eq:eachNzz},
leading to discrepancies between the two methods.
They represent different physics in this case.
It is worth noting again that this discrepancy can not be eliminated as cell size approaches zero.

For sufficiently smooth magnetization, the convergence speed estimated by Eqs.~\eqref{eq:nearandfar} and Eqs.~\eqref{eq:NearFarAnother} are not faster than $h$. However, numerical experiments indicate a convergence speed of $h^2$ for sufficiently smooth magnetization. Therefore, the lower bound of the convergence speed is much worse than that observed in numerical experiments.
To achieve a better lower bound estimation for the convergence speed, one may need methods beyond the simple near and far field splitting, and cell-by-cell analysis. Considering the potential for error cancellation among cells, it is crucial to carefully analyze the correlations between errors of different cells.

The section of our proof that demonstrates the near field is zero for uniform magnetization closely resembles the approach described in literature \cite{jackson2021classical}. However, it serves a different purpose: while the literature \cite{jackson2021classical} aims to establish the relationship between molecular polarizability and electric susceptibility, our goal is to prove that different magnetization tensors yield consistent results for both non-uniform and uniform magnetizations.

The conclusion does not readily extend to two-dimensional materials in three-dimensional space.
The magnetization of a two-dimensional thin material that is one cell thick cannot be considered a three-dimensional H\"{o}lder continuous function.
Specifically, the symmetry between $x$, $y$, and $z$ is broken for two-dimensional materials. 

\section{Conclusion}
\label{sec:conc}

We prove that the \emph{UMC}, \emph{UMCD}, and \emph{dipole} methods, their asymptotic expansions, and the Cauchy principal value method yield consistent results in three-dimensional space as the cell size approaches zero,
namely,
\begin{equation}
    \lim_{h \to 0} \bm{H}_\text{d}(\bm{r})
    =
    \lim_{h \to 0} \bm{H}_\text{c}(\bm{r})
    =
    \lim_{h \to 0} \bm{H}_\text{cd}(\bm{r})
    =
    \bm{H}_\text{p}.
\end{equation}
If the self-contribution of the cell to the demagnetization field is included in both the \emph{UMC} method and the \emph{UMCD}, then
\begin{equation}
\begin{aligned}
    &\lim_{h \to 0} \bm{H}_\text{d}(\bm{r}) - \frac{1}{3}\bm{M} \\
    =~ &\lim_{h \to 0} \bm{H}_\text{c}^\text{full}(\bm{r}) = \lim_{h \to 0} \bm{H}_\text{cd}^\text{full}(\bm{r}) \\
    =~ &\bm{H}_\text{p} - \frac{1}{3}\bm{M} = \bm{H}.
\end{aligned}
\end{equation}
The primary challenge of the proof arises from the fact that the magnetization field contributed by nearby cells remains finite as the cell size approaches zero, as do the differences in these methods.
To address this challenge, we utilize the fact that the contribution from all cells in a simple cubic volume to the demagnetization field is zero for uniform magnetization. 
This allows us to subtract a uniform magnetization from the total magnetization, leaving a zero magnetization at the points where the field is being evaluated. 
Thus, the values of the demagnetization tensor for short distances do not affect the total demagnetization field.
Additionally, we provide a lower bound on the convergence speed. The conclusions are validated by numerical experiments. 
We acknowledge that while the conclusions are applicable in three-dimensional materials, further investigation is required for two-dimensional materials in three-dimensional space.

The agreement among these demagnetization tensors is significant.
The principal integral method offers a macroscopic perspective, while the demagnetization tensor on a grid provides a microscopic perspective. As the cell size approaches zero, results from different methods converge to the same limit, bridging the macroscopic and microscopic perspectives and making them comparable. 
This agreement shows the correctness of the numerical methods and that different microscopic physics models are equivalent in the limit $h \to 0$. The so-called method error for the \emph{dipole} method can be attributed to a type of discrete error. 

In the calculation of the demagnetization tensor for the \emph{UMC} method, both numerical integrals and lengthy exact analytical formulas introduce complexity in implementation. Additionally, exact analytical formulas can suffer from numerical stability issues, such as catastrophic cancellation \cite{Donahue2007}. In contrast, the implementation of the demagnetization tensor in the \emph{dipole} method is considerably simpler. For numerical calculations, the consistency among these methods allows us to employ the \emph{dipole} method in situations where convergence speed is not a critical factor.

\begin{acknowledgments}
This work was in part supported by NIH grant R01 EB 031078. The content is solely the responsibility of the authors and does not necessarily represent the official views of the National Institutes of Health.

\end{acknowledgments}

\section{AUTHOR DECLARATIONS}

\subsection{Conflict of Interest}
The authors have no conflicts to disclose.

\subsection{Author Contributions}
Hao Liang: Conceptualization (equal); Methodology (lead); Software (lead); Validation (equal); Formal analysis (lead); Investigation (equal);
Writing - original draft (lead); Writing - review \& editing (equal); Visualization  (lead).
Xinqiang Yan: Conceptualization (equal); Validation (equal); Investigation (equal); Resources (lead); Writing - review \& editing (equal); Funding acquisition (lead); Supervision (lead).

\section{DATA AVAILABILITY}
The data that support the findings of this study are available from the corresponding authors upon reasonable request.

\appendix
\section{Solution for problem I}
\label{sec:app1}
$H_z$ at $\bm{r} = 0$ for problem I is given by
\begin{widetext}
\begin{equation}
\begin{aligned}
&\frac{1} {12 \pi  (\alpha +1)^2 (\alpha +2)} \bigg[ -4 \pi  \alpha ^2-18 \alpha ^2 \log (2)-24 \alpha ^2 \log \left(2-\sqrt{3}\right) +36 \alpha ^2 \log \left(\sqrt{3}-1\right) -9 \alpha ^2 \psi ^{(0)}\left(\frac{\alpha +1}{4}\right)\\
&~~~~ +9 \alpha ^2 \psi ^{(0)}\left(\frac{\alpha +3}{4}\right)-8 \pi  \alpha -36 \alpha -54 \alpha  \log (2) -72 \alpha  \log \left(2-\sqrt{3}\right)+108 \alpha  \log \left(\sqrt{3}-1\right) \\
&~~~~ -27 \alpha  \psi ^{(0)}\left(\frac{\alpha +1}{4}\right)+27 \alpha  \psi ^{(0)}\left(\frac{\alpha +3}{4}\right) -18 \psi ^{(0)}\left(\frac{\alpha +1}{4}\right) +18 \psi ^{(0)}\left(\frac{\alpha +3}{4}\right)\\
&~~~~ -24 \sqrt{2} \alpha ^2 F_1\left(\frac{\alpha +1}{2};-\frac{1}{2},1;\frac{\alpha +3}{2};-\frac{1}{2},-1\right) -36 \sqrt{2} \alpha  F_1\left(\frac{\alpha +1}{2};-\frac{1}{2},1;\frac{\alpha +3}{2};-\frac{1}{2},-1\right) \\
&~~~~ -24 \sqrt{2} F_1\left(\frac{\alpha +1}{2};-\frac{1}{2},1;\frac{\alpha +3}{2};-\frac{1}{2},-1\right) +18 \sqrt{2} \alpha ^2 \, _2F_1\left(\frac{1}{2},\frac{\alpha +1}{2};\frac{\alpha +3}{2};-\frac{1}{2}\right)\\
&~~~~ +18 \sqrt{2} \alpha  \, _2F_1\left(\frac{1}{2},\frac{\alpha +1}{2};\frac{\alpha +3}{2};-\frac{1}{2}\right)  +12 \sqrt{2} \, _2F_1\left(\frac{1}{2},\frac{\alpha +1}{2};\frac{\alpha +3}{2};-\frac{1}{2}\right)\\
&~~~~ +18 \alpha ^2 \Phi \left(-1,1,\frac{\alpha +3}{2}\right)+54 \alpha  \Phi \left(-1,1,\frac{\alpha +3}{2}\right)  +36 \Phi \left(-1,1,\frac{\alpha +3}{2}\right)-4 \pi -72-36 \log (2)
\\
&~~~~ -48 \log \left(2-\sqrt{3}\right)+72 \log \left(\sqrt{3}-1
\right)
\bigg],
\end{aligned}
\end{equation}
\end{widetext}
where log is the natural logarithm, $\psi^{(0)}$  the polygamma function of order 0,
$F_1$ the Appell series,
$_2F_1$ the hypergeometric function,
and $\Phi$ the Lerch transcendent.

\bibliography{references}

\end{document}